\documentclass[12pt]{article}
\usepackage{amssymb}
\title{The origin of the difference between space and time }
\author{Hrvoje Nikoli\'c  \\
Theoretical Physics Division, Rudjer Bo\v{s}kovi\'{c} Institute, \\
P.O.B. 180, HR-10002 Zagreb, Croatia \\
{\normalsize e-mail: hnikolic@irb.hr} \\
\makebox[1in]{} \\
}
\date{\today}
\begin{document}
\maketitle
\begin{abstract}
All differences between the role of space and time in nature are 
explained by proposing the  
principles in which none of the spacetime
coordinates has an {\it a priori} special role. Spacetime 
is treated as a non-dynamical
manifold, with a fixed global $\mathbb{R}^D$ topology. 
Dynamical theory of
gravity determines only the metric tensor on a fixed manifold. 
All dynamics is treated as a Cauchy problem, so   
it {\em follows} that one coordinate takes a special role.                       
It is proposed that {\em any} boundary condition that is 
finite everywhere leads to  
a solution which is also finite everywhere. This explains 
the $(1,D-1)$ signature of the metric, 
the boundedness of energy from below, the absence of tachyons, and other related
properties of nature. The time arrow is explained by proposing that
the boundary condition should be ordered. 
The quantization is considered as a boundary condition for field
operators.
Only the physical degrees of freedom are quantized. 
\end{abstract}
\vspace*{0.5cm}

\section{Introduction}

One of the most fundamental principles of modern theoretical physics is
the principle of Lorentz covariance. This principle  essentially says that 
all fundamental physical theories should treat space and time coordinates in
the same way, up to a negative relative sign in the metric of 
spacetime.
However, it is known that space and time coordinates are not really
treated in the same way, 
and that these different treatments cannot be explained only from  
the negative relative sign in the metric. One has to introduce some
additional principles in order to explain and describe the observed
different roles of space and time in nature. Let us make a list of some
very known principles and observational facts that explicitly state
that space and time should be treated in different ways:

\vspace{0.5 cm}

-- There are a few space coordinates, but there is precisely one time 
coordinate.

-- There is a time arrow, but there is nothing like a space arrow.

-- Psychologically, we experience time and space in completely different
ways; we remember the past and not the future, which refers to time, not to
space.

-- We can travel in space in all directions, but we cannot do that in time.

-- The entropy grows with time, but not with space.

-- There is a causality principle, which refers to time, not to space; in
classical electrodynamics, one uses only retarded solutions and disregards
advanced solutions, which again refers to the sign of time, not that of space.

-- The separation of causally connected events should be timelike or
light-like, but cannot be spacelike; the 4-momentum of a physical particle
should be timelike or light-like, but cannot be spacelike.

-- Time has a special role in the canonical (i.e., Hamiltonian) formalism; in
field theory (of real scalar fields, for simplicity), the set of all
degrees
of freedom is given by all space points ${\bf x}$, not by all space-time
points $({\bf x},t)$; in order to quantize fields, we propose equal-time
(anti)commutators, not equal-space (anti)commutators; field operators
(anti)commute for spacelike separations, not for timelike separations.

-- The time component of the 4-momentum (energy) must be positive (or zero),   
while the space components of the 4-momentum can have both signs; the quantum
operator of the space inversion is unitary, while the quantum
operator of the time inversion is anti-unitary.

-- In the quantum theory of particles (i.e., first
quantization) there is an $\hat{{\bf x}}$-operator, but there is no  
$\hat{t}$-operator.

\vspace{0.5 cm}

If one believes that the fundamental laws of nature should possess a 
certain simplicity and symmetry, then it is reasonable to believe 
that the fundamental laws 
should have such a form that none of the space-time coordinates has
an {\it a priori} special role. If this is so,
none of the itemized laws can be fundamental. From
some more fundamental laws it should rather follow that one
of the coordinates {\em must} take a special role, by a mechanism which
can be viewed as some kind of spontaneous symmetry breaking.

The idea that the different roles of space and time are consequences of
spontaneous symmetry breaking is not new. In \cite{ivan,moffat,sardan}   
the possibility is considered that this is achieved via the Higgs mechanism.
The aim of this article is to give a proposal for a different mechanism  
which gives different roles to space and time, a mechanism which does not  
require the introduction of the Higgs field. 

It is a tradition among almost all physicists that
only finding the correct equations of motion is regarded as a really 
fundamental task, while the question of the boundary 
conditions is regarded as a  
secondary problem. Here I leave such a viewpoint. 
I consider the
question of the boundary conditions as an equally fundamental question 
as the question of
the equations of motion themselves. Therefore, I postulate some
principles which the boundary condition of the Universe should obey. These
principles I choose in such a way that none of space-time coordinates has
an {\it a priori} 
special role, but that they can still explain the known differences of the   
role of space and time in the Universe. 

The difference between space and time emerges from the viewpoint that nature  
{\em must} choose some $(D-1)$-dimensional sub-manifold on which the boundary 
condition will be imposed. This automatically gives a special status to 
one particular coordinate, the coordinate which is 
constant on this sub-manifold. This {\em
is} the mechanism of spontaneous symmetry breaking in my approach. 
I propose essentially three additional principles. First,  
spacetime is a non-dynamical manifold, with a fixed global $\mathbb{R}^D$
topology. Dynamical theory of gravity determines only the metric tensor on it.
Second, I propose that {\em any} boundary condition which is 
finite everywhere leads to 
the solution which is also finite everywhere. This explains the
hyperbolicity, i.e., $(1,D-1)$ signature of the metric. (It is interesting to
note that there is an attempt to explain the hyperbolicity by certain
anthropic arguments \cite{teg}. My approach is based on the same mathematical 
properties of hyperbolic and non-hyperbolic equations exploited 
in that work, but I choose different arguments to favor  
hyperbolic equations only.)
This second principle also explains 
the boundedness of energy from below, the absence of tachyons, and other related
properties of nature. The third principle states that the boundary condition
is ordered, rather than random. It explains the time arrow.
The quantization is considered as a boundary condition for the field
operators.
Only physical degrees of freedom are quantized. This, together with the     
treatment of spacetime as a non-dynamical background, resolves the problem    
of time in quantum gravity, at least at the conceptual level. Possible    
paradoxes connected with the possibility of time travel are excluded by   
my choice of topology. 

In Sec.~\ref{SEC2} I present the 
main physical and mathematical ideas which led me to find the principles    
which can describe the nature of space and time and explain the differences
between them. 
In Sec.~\ref{SEC3} I give a precise formulation of these
principles, as a set of axioms which classical physics should obey. 
The purpose of Sec.~\ref{SEC4} is to discuss in more detail 
the origin of various differences between
the role of space and time in classical physics, 
emphasizing that they all emerge from the axioms 
of Sec.~\ref{SEC3}. 
In Sec.~\ref{SEC5} I discuss the origin of the difference
between the role of space and time in quantum physics. 
The connection with
classical physics is the most manifest in the Heisenberg picture, which
I use to formulate the quantization as a boundary condition for field
operators. 
In Sec.~\ref{SEC6} I discuss whether the second principle that I
propose is satisfied for the known physical theories and 
what new consequences can emerge from this principle. 
In Sec.~\ref{SEC7} I discuss whether 
some of my axioms can be rejected or weakened. In addition, I make  
some remarks on the question of dimensionality of space. 
Sec.~\ref{SEC8} is devoted to concluding remarks. 

\section{The main ideas}
\label{SEC2}

In this section I give the main 
physical and mathematical ideas which led me
to find the principles proposed in Sec.~\ref{SEC3}.  
Sec.~\ref{SEC2} is intended 
to be very pedagogical, but not too exhaustive. It is also intended to be
intuitive, rather than rigorous.   

Let us start from the origin of the time arrow. Most of physicists agree 
that all manifestations of the time arrow (except the arrow connected with 
the direction of the expansion of the Universe) are consequences of the 
thermodynamic time arrow, i.e., of the fact that disorder increases with 
time. 
The fact that disorder grows with time is equivalent 
to the statement that the Universe was quite ordered in the past. Thus,
the only real problem with the time arrow is to explain {\em why}    
the Universe was so ordered at some instant of time of its 
evolution. Since I cannot find any 
convincing explanation of this (except the anthropic principle
\cite{hawk2}), I shall
take this as one of my fundamental postulates. It is enough to postulate  
that at some ``initial" instant of time (not necessarily to be 
the earliest instant) all fields and matter must be in 
some partially ordered configuration in all space regions, 
but in such a way that ``initial" velocities 
have random space-directions. I require random directions of velocities 
because then {\em both time directions are equivalent}, in the sense that 
disorder increases in both directions from this ``initial" instant. 
The present velocities are obviously not random, since they lead to the
increasing order in the negative time direction.
The ``initial" instant is
actually the instant of minimal entropy. 

The next question considered is why is {\em time} the 
coordinate which takes a special role? Why is this not the
$z$-coordinate,  
for example? Or why is there no more than one coordinate which takes the 
role similar to that of time? The answer to this question can be easily 
found if one treats the dynamics of the Universe as a Cauchy problem. 
To solve a partial differential equation in $D$ dimensions, one first needs 
to fix some $(D-1)$-dimensional sub-manifold (Cauchy surface) 
on which the Cauchy data  
will be imposed. This automatically gives a special status 
to one particular coordinate, the coordinate which is constant on 
this sub-manifold. If we, in addition, require that the 
differential equation should provide a stable evolution of the Cauchy data, 
then for a second-order differential equation two necessary conditions 
must be fulfilled \cite{teg}: First, the equation must be hyperbolic, 
which corresponds to the $(1,D-1)$ signature of the metric. Second, 
the Cauchy surface must be spacelike, i.e., the boundary condition 
must be the initial condition.  

Let us illustrate this on a free-field equation
\begin{equation}\label{1}
 \partial_{\mu}\partial^{\mu}\phi (x) +m^2 \phi (x)=0 \; .
\end{equation}
All known free fields satisfy this equation, including the Dirac field too. 
We assume that the metric has the form $g_{\mu\nu}={\rm diag}(1,-1,-1,-1)$. 
$m^2$ is some real parameter that can be positive, zero, or negative. If we
are looking for the solution of the form $\phi (x)=\exp (ik\cdot x)$, we 
find the dispersion relation 
\begin{equation}\label{2}
 k_0^2 - {\bf k}^2 =m^2 \; .
\end{equation}
In general, any component of $k$ can be complex. However, if we require that 
the solution is finite for any value of $x$, including the cases when
some of the components of $x$ is $\pm \infty$, we conclude that all 
components of $k$ must be real. Now let us suppose that $m^2 >0$. In this   
case, the real vector ${\bf k}$ can be arbitrary, since then $k_0$ is also
real. However, $k_0$ cannot take an arbitrary real value, but must
rather    
satisfy $k_0^2 \geq m^2$. We can construct the general solution of
(\ref{1}) which is finite everywhere as a Fourier expansion over plane-wave
solutions. For some fixed $t$, it can have an {\em arbitrary} 
(finite everywhere)
dependence on ${\bf x}$. However, for fixed $z$, for example, it cannot
have an arbitrary time-dependence, because the spectrum of $k_0$ is truncated. 
Therefore, if we require that the {\em arbitrary} boundary 
condition that is finite everywhere leads to the solution which is also 
finite everywhere, then the boundary condition must be the initial
condition.  

Using a similar argument one can also see that $m^2$ cannot be negative, 
because otherwise, owing to the fact that there is more than one space coordinate, 
no 3 components of $k$ could take arbitrary real values, without 
leading to the imaginary fourth component of $k$. 

One can also 
easily generalize the analysis to the flat metric with the $(n,m)$ 
signature, and conclude that the arbitrary boundary condition 
that is everywhere
finite can lead to the solution which is also finite everywhere only if 
$n$ or $m$ is equal to 1. 

Now we can already see the main idea why one coordinate, so-called time,
takes a special role. The dynamics is described by some partial differential 
equations in $D$ dimensions that treat all coordinates in the same way,  
up to some signs in the metric, which can generally take the $(n,m)$
signature $(n+m=D)$. Thus the differential equations  are covariant with
respect to the SO($n,m$) group of coordinate transformations.
However, the differential equations 
do not describe the dynamics uniquely; one must also fix
some boundary condition. To do that, one first must fix the boundary
itself,
which is some ($D-1$)-dimensional sub-manifold. This defines the remaining one
coordinate which has the same value at the whole ($D-1$)-dimensional subspace.
By imposing that the {\em arbitrary}
finite everywhere boundary condition leads to a solution which is also 
finite everywhere, we obtain that all coordinates on this boundary must
have the same sign of the metric and that the remaining one coordinate must
have the opposite sign of the metric. Thus we {\em derive} the Lorentz 
invariance SO($1,D-1$) (isomorphic to SO($D-1,1$)).
This also leads to some constraints on the form of the differential
equations,
including the sign of $m^2$. In addition, we impose that the boundary      
condition must be {\em ordered} in a described sense, from which we
{\em derive} the second law of thermodynamics and thus the causal role of
the time coordinate. 
  
I also want to clarify some   
conceptual details that are important for a deeper understanding of gravity.
Physicists are used to think that there is a great 
difference between the gravitational field and all other fields, because
other
fields describe some dynamics for which spacetime serves as a background,
while the gravity field describes the dynamics of spacetime itself. So,   
they often imagine that spacetime itself cannot exist 
without the existence of the gravity field $g_{\mu\nu}(x)$, whereas it can 
exist without other fields (which corresponds to $T_{\mu\nu}=0$ in 
the Einstein equation), 
and without the dark energy (which can be absorbed into a term contributing 
to the total $T_{\mu\nu}$).
However, a manifold with coordinates $x^{\mu}$ can 
be well defined even without the metric being defined. This leads to 
the possibility of interpreting the gravitational field 
in such a way that it differs much less from the other fields. 
Such an interpretation could be useful in order to formulate a consistent
theory of quantum gravity. 

When solving the Einstein equation, one can forget that $g_{\mu\nu}(x)$
represents the metric tensor; it can be viewed just as some second-rank
tensor field. Moreover, solving the Einstein equation as a Cauchy problem
requires that the topology
of spacetime should be fixed before the actual solving. More 
precisely, the Cauchy problem is well posed only if the topology takes the form 
$\Sigma\times \mathbb{R}$ on the {\em global} level, where $\Sigma$ represents  
the topology of the Cauchy surface. (Note that, in practice, the Einstein
equation is
usually not solved as a Cauchy problem; it is solved by imposing some
symmetry conditions of the metric on the {\em whole} spacetime. The various
solutions satisfying these conditions are then recognized as representing   
various topologies.) In this article I propose that the whole dynamics
{\em should} be treated as a Cauchy problem, so I propose that the
topology of spacetime is not a dynamical entity. 
Since I require that none of space-time coordinates should have an {\it a
priori} special role, the topology should also be symmetrical in that sense.
Therefore, I choose $\mathbb{R}^D$ as a {\em global} topology. Note finally that
the
condition $D=4$, as well as the (1,3) signature of the metric,
must also be imposed by the initial condition in the
Cauchy-problem approach to the Einstein equation.

\section{The formulation of principles}
\label{SEC3}

In this section the precise formulation of principles that I propose 
is given as a set of axioms. 
These axioms refer mainly to classical physics, while
the transition to quantum physics is discussed in Sec.~\ref{SEC5}.

\newtheorem{axiom}{Axiom}
\newtheorem{definition}{Definition}

\begin{axiom}\label{ax1}
 There exists a manifold ${\cal M}$ which can be {\bf globally} bijectively
 mapped to
 the set $\mathbb{R}^D$, where $D$ is a fixed positive integer.
\end{axiom}
This axiom says that spacetime is continuous, $D$-dimensional, infinite, 
and
predynamical. The mapping in Axiom \ref{ax1} defines the coordinates $x\equiv
\{x^1,
\ldots ,x^D\}\in \mathbb{R}^D$. Next we introduce a metric tensor on ${\cal
M}$ which is 
a symmetric second-rank tensor which must satisfy the following axiom.
\begin{axiom}\label{ax2}
 For each point $x$ there exists a neighborhood $U$, non-negative
 integers $n$, $m$ satisfying $n+m=D$, and coordinates, such that   
 the metric tensor possesses $n$ positive and $m$ negative eigenvalues
 on $U$.
\end{axiom}
This axiom says that for each point there exist numbers $n,m$ 
and coordinates such that the metric 
is invariant with respect to SO$(n,m)$ coordinate 
transformations at this point. This is a generalization of the Lorentz
SO$(1,3)$
invariance. It is also   
important to note that from Axiom \ref{ax2} it follows that the 
metric possesses the {\em global} decomposition into $D=n+m$, i.e., if, 
for example, 
the manifold ${\cal M}$ has the (1,3) signature of the metric
at some point, then it has the same signature on the whole ${\cal M}$.

Now we introduce  
dynamics, described by some fields $\varphi_{a}(x)$. The metric 
tensor can also be one of dynamical fields, but this is not necessary. 
For dynamical fields we require the following axiom: 
\begin{axiom}\label{ax3}
 Dynamical fields satisfy partial differential equations (with derivatives
 with respect to $x^{\mu}$) and for each point $x$ there exist
 coordinates such that the equations are covariant 
 with respect to SO$(n,m)$ coordinate transformations at this point, 
 where $n,m$ are determined by Axiom \ref{ax2}.
\end{axiom}
To construct such differential equations, we do not usually have to
worry
about the precise values of $n$ and $m$, since these equations look formally
the same for various $n,m$ when written in a manifestly covariant form. 
Use of Lagrangian techniques further simplifies the construction
of such equations.

The knowledge of the differential equations does not determine
dynamical fields uniquely. We want to understand the principles 
which nature obeys in order to pick up a particular solution 
that corresponds to the actual Universe.    
Now the essence of my philosophy is as follows: It is redundant for 
nature to choose some differential equations {\em and} some particular
solution.
Nature actually chooses some differential equations and some
boundary condition. The crucial point is that nature {\em must} first 
choose some $(D-1)$-dimensional sub-manifold ${\cal M}_{B}\subset {\cal M}$  
on which the boundary condition will be imposed,   
so nature really {\em does} choose
it. This choice is not considered as a mathematical
convenience, but rather as a real event in nature. Such a viewpoint can    
look slightly metaphysical, but we shall see that such a viewpoint leads 
to a natural explanation of the known differences between the roles of
space and time, as well as to some new predictions. Furthermore, we
shall see that, for a given universe, this ``canonical" 
sub-manifold ${\cal M}_{B}$ can be uniquely identified, at least in
principle. In the following 
I propose some axioms that refer to the properties of 
this ``canonical" ${\cal M}_{B}$
and the corresponding boundary condition, which nature should obey.
     
If the differential equations are of the $k$-th order in the field
$\varphi_{a}(x)$, then it is convenient to choose some connected boundary
${\cal M}_{B}$ and to fix $\varphi_{a}(x)$ and all its normal derivatives on
it, up to 
the $(k-1)$-th derivative. If this is done for all fields appearing
in the differential equations, the Cauchy-Kowalevska theorem provides
that the solution is then unique. (Strictly speaking, this theorem
also requires the analyticity of the boundary condition and provides 
the analyticity of the solution. However, I shall assume that the
Cauchy problem is well posed also for smooth enough boundary conditions 
which are not necessarily analytic.) 
Therefore, I propose the following axiom:  
\begin{axiom}\label{ax4}
 The boundary ${\cal M}_{B}$ is a connected $(D-1)$-dimensional sub-manifold
 which can be {\bf globally} bijectively mapped to the set $\mathbb{R}^{D-1}$.
\end{axiom}
It is understood 
that the boundary condition fixes the fields $\varphi_{a}(x)$ and all its
normal derivatives up to the $(k-1)$-th derivative. Thus, because of  
Axiom \ref{ax1}, the topology 
of ${\cal M}_{B}$ proposed in Axiom \ref{ax4} 
is the only one that can lead to a well-posed boundary-condition problem.   

Let us now introduce the following definition:
\begin{definition}\label{def1}
 A function $\varphi :X\rightarrow \mathbb{C}$ is {\bf regular} on a  
 domain $X\subseteq {\cal M}$ if $|\varphi(x)|$ is bounded from above 
 for every $x\in X$.
\end{definition}
In other words, a regular function is a function which is finite everywhere.
It is quite reasonable to require that physical fields should be regular.
However,
it is known that some fields, such as the metric tensor $g_{\mu\nu}$, the 
connection
$\Gamma^{\rho}_{\mu\nu}$, and the vector potential $A_{\mu}$ do not have to be
regular. I shall refer to such fields as gauge fields. Only physical
fields, such as the scalar curvature $R$ and the field 
strength $F_{\mu\nu}$ have to
be regular, whereas the gauge fields can possess only such irregularities which
do not lead to irregularities of the corresponding physical fields. Having
this in mind, I introduce the following definition:
\begin{definition}\label{def2} 
 The field $\varphi_a$ is {\bf essentially regular} on $X$ if its
corresponding
 physical field is regular on $X$. The 
 metric field is essentially regular on 
 $X$ if it is essentially regular as a field and satisfies Axiom \ref{ax2}    
 on $X$.
\end{definition}
I have no intention to give a rigorous definition of a physical field.
Let me just note that fields appearing in the Lagrangians which do not
possess any kind of gauge symmetry are its own physical fields.    

Now we are ready to propose the following axioms:
\begin{axiom}\label{ax5}
 For a given signature of the metric there exists ${\cal M}_B$ such that 
 {\bf every} boundary condition                                            
 essentially regular on ${\cal M}_B$ leads
 to the solution essentially regular on ${\cal M}$.
\end{axiom}
\begin{axiom}\label{ax6}
 The Cauchy surface ${\cal M}_B$ is chosen in such a way as to satisfy the
requirement 
 of Axiom \ref{ax5}. The boundary condition is essentially regular on
 ${\cal M}_B$. 
\end{axiom}
Axiom \ref{ax5} is central and the most important axiom of this
article. This is actually not the constraint on the boundary condition, but
rather on the signature and on the    
possible forms of the dynamics, i.e., on the possible
equations of motion.
As we shall see, this Axiom explains the hyperbolicity of the equations of
motion, i.e., the $(1,D-1)$ signature of the metric. It also explains a lot
of known differences between the role of space and time itemized in the    
Introduction. Finally, it leads to some new predictions. All that will be
discussed in later sections. Here I want to explain that axioms of this
section lead to a new philosophy of the logical (not temporal) order which
nature must follow when it chooses the conditions which uniquely determine
the Universe. 

Nature first chooses the dimension $D$ of the manifold
${\cal M}$, according to Axiom \ref{ax1}. 
(Of course, the word ``chooses'' should not be understood in the anthropomorphic sense.)
Then it chooses the signature
$(n,m)$
according to Axiom \ref{ax5}. After that it chooses ${\cal M}_B$
according to
Axiom \ref{ax6}. 
The next step is to choose a set of
fields $\{\varphi_a \}$ which will describe the dynamics, making a
difference between 
the physical and the gauge fields, but not yet specifying its specific
dependence on $x$. The very next step, the central one in my
philosophy, is to choose 
the differential equations (or Lagrangian) which will
provide that {\em any}
essentially regular boundary condition will lead to an essentially regular
solution, according to Axiom \ref{ax5}. Of course, these differential
equations must also satisfy some
additional principles, such as covariance (Axiom \ref{ax3}) and probably  
some other principles, which are not important here. At the end it only
remains to choose some particular boundary condition on ${\cal M}_B$,      
according to Axiom \ref{ax6}, which
then uniquely determines the Universe.
   
The six axioms, proposed so far, still cannot explain all the differences
between the role of space and time itemized in the Introduction. We need
one additional axiom which will provide that disorder increases with 
time and thus explains the time arrow.
This axiom must essentially say that the boundary condition is not
completely random, but rather ordered somehow, as discussed in Section
2. It is not easy to formulate this axiom in a mathematically rigorous
way. Thus I formulate this in a way which is not very rigorous, but
rather intuitive:  
\begin{axiom}\label{ax7}
 The boundary condition on ${\cal M}_B$ is  
 partially ordered, rather than random. 
 In particular, absolute values of various fields are not
 homogeneous, but rather lumped in localized lumps. However, the field
 derivatives in the normal direction to ${\cal M}_B$, needed for the
 uniqueness
 of the solution of the Cauchy problem, are random. 
\end{axiom}  
The last sentence in Axiom \ref{ax7} corresponds to the assumption    
that the initial velocities are random, which provides that disorder  
increases in both time directions from the so-called initial hypersurface  
${\cal M}_B$, so both time directions are equivalent.    

\section{The differences between space and time in classical physics}
\label{SEC4}

In this section I discuss how all the differences between
the
role of space and time in classical physics emerge from the axioms of 
Section 3. However, it is important to note that the most of the discussion
is valid even if novel principles of this article are not realized
in nature. 
Only Axioms \ref{ax1} and \ref{ax5} are really novel principles, in the
sense that they differ from the conventional point of view and can be
tested, at least in principle. 
In order to provide a complete and clear picture, I find it 
necessary to review some already known results.

If dynamical fields satisfy second-order differential equations, 
then it follows from Axioms \ref{ax3} and \ref{ax5} that the 
differential equations must be hyperbolic, i.e., the signature of the 
metric must be $(1,D-1)$ (see, for example, 
\cite{teg} and references therein). 
We shall see in Sec.~\ref{SEC6} that Axiom \ref{ax5} also explains  
why dynamical fields are not described by differential equations 
of order higher than second. 

In order to satisfy Axiom \ref{ax5}, the hyperbolicity is necessary, 
but not sufficient. For free fields, for example, we have seen in 
Section 2 that $m^2$ cannot be negative.
The fact that $m^2$ cannot be negative explains why the $D$-momentum of a  
free physical particle cannot be spacelike. If we assume that the
propagation velocity of a free wave packet is given by the so-called 
group velocity 
\begin{equation}\label{7}
 {\bf v}_g =\frac{d\omega}{d {\bf k}} \; , 
\end{equation} 
where $\omega =\sqrt{{\bf k}^2 +m^2}$,  
then we see that there are no velocities greater than $c\equiv 1$, 
which then explains why the separation of causally connected events cannot
be spacelike, at least for the free case.  

However, it is fair to mention that the propagation velocity of a 
free wave packet is not always given by (\ref{7}). Thus it is not strange
that there are solutions of all known free relativistic wave equations
(such as free Klein-Gordon, Maxwell, and Dirac equations) which propagate 
with superluminal velocities, i.e., velocities that are greater than $c$
\cite{superlum}. However, there are no real paradoxes with these solutions
because it turns out that the corresponding physical quantities (such as the    
Poynting vector for the electromagnetic field) do not propagate faster  
than $c$. Thus the principle that no energy or information can propagate faster
than $c$ is not violated, and this is what we understand when we claim that 
fields do not propagate faster than $c$.  

The requirement that $m^2 \geq 0$, which was 
obtained for free fields, in the case of interacting fields 
generalizes to the requirement that the energy should be bounded from 
below. To see this, we 
consider the Lagrangian for the real scalar field $\phi (x)$:  
\begin{equation}\label{8}
 {\cal{L}}=\frac{1}{2}\partial_{\mu}\phi\partial^{\mu}\phi -V(\phi ) \; ,
\end{equation}
where
\begin{equation}\label{9}
 V(\phi )=-\frac{\mu^2}{2}\phi^2 +\frac{\lambda}{4}\phi^4 \; .
\end{equation}
The parameters $\mu^2$ and $\lambda$ are real constants. The corresponding
equation of motion can be written in the form
\begin{equation}\label{10}
 \partial_{\mu}\partial^{\mu}\phi (x) +m_{\rm eff}^2 (x) \phi (x)=0 \; , 
\end{equation}
where 
\begin{equation}\label{11}
 m_{\rm eff}^2 (x)=-\mu^2 +\lambda \phi^2 (x) \; .
\end{equation}
If we require the stable time evolution, and 
if $\lambda\neq 0$, then for a large $\phi^2(x)$ 
the relation $m_{\rm eff}^2 (x) \geq 0$ must be fulfilled
\cite{rajag}. This means
that the relation $\lambda >0$ must be fulfilled, which is actually 
the consequence of Axiom \ref{ax5}, because the stability requirement is
essentially the same requirement as
Axiom \ref{ax5}. 

Let us now see what it has to do with the sign of 
energy. We introduce the canonical energy-momentum tensor
\begin{equation}\label{14}
 \Theta^{\mu}_{\nu} =\frac{\partial{\cal{L}}}{\partial (\partial_{\mu}\phi)}
 \partial_{\nu}\phi -g^{\mu}_{\nu} {\cal{L}} \; .
\end{equation}
The corresponding energy-density for the Lagrangian (\ref{8}) is
\begin{equation}\label{15}
 {\cal{H}}=\Theta^0_0=\Theta^{00}=
 \left[ \frac{\dot{\phi}^2}{2} + \frac{(\nabla\phi)^2}{2} \right]
 + V(\phi) \; .
\end{equation}  
The term in the square bracket represents the kinetic part of $\Theta^0_0$.
We see that it has the definite (positive) sign. It is easy to see that
owing 
to the $(1,D-1)$ signature of the metric, no other component of 
$\Theta^{\mu}_{\nu}$ or $\Theta^{\mu\nu}$ has definite sign of its
kinetic part for $D >2$. Since $\lambda >0$, 
we see that $V(\phi)$ is bounded from
below. Thus we see that the boundedness of the energy from below is 
actually the consequence of Axiom \ref{ax5} (i.e., the stability  
requirement). 
A similar connection between Axiom \ref{ax5} and 
the boundedness of the energy from below can be seen in a similar way    
for most of other Lagrangians. The positivity of energy is 
then obtained from the appropriate energy shift, which does not change
the physical laws (except gravity, at least in the conventional
approach). 

We see that if $E$ is some admissible energy, then $-E$ may not be
admissible energy. The consequence of this is that energy does not transform
as a time component of a $D$-vector with respect to time inversion.

Let us discuss now the consequences of Axiom \ref{ax7}. 
(In the following I use the term ``disorder"
rather than ``entropy", because the former is a more general concept, 
while the
latter corresponds to some particular measure of disorder, which can be 
inappropriate for some purposes). 
According to this axiom, the so-called 
initial state of the Universe is quite ordered. We assume that the
degree of orderliness is homogeneous on the initial
spacelike manifold ${\cal M}_B$. This manifold defines the natural
foliation of
spacetime into the class of spacelike manifolds $\Sigma (t)$, with the
property $\Sigma (0) ={\cal M}_B$. We choose $t$ in such a way that the
orderliness is homogeneous (at least at some large scale) on the whole
$\Sigma$ for any fixed $t$. Disorder increases in both time directions
from $t=0$, so there is no a 
special time direction. The causal, psychological, and electrodynamic 
time arrows are consequences of the disorder increase. 
The positive time direction is defined as a direction
from ${\cal M}_B$ to the present time.
Thus $t$ defines the natural cosmological
time, but still not uniquely, because $t$ can be replaced by some $h(t)$,
where
$h$ is some strictly increasing function, satisfying $h(0)=0$. In order to
define the time coordinate uniquely, we can require $g_{tt}=1$.
For a given universe, ${\cal M}_B$ can be uniquely identified as a
$(D-1)$-dimensional 
spacelike manifold $\Sigma$ with the smallest measure
of disorder (entropy). The instant $t=0$ can be considered as the instant of
the ``creation" of the Universe (whatever this means) by some yet
unknown mechanism.

According to Axiom \ref{ax7}, the fields are initially lumped. Since 
fields cannot propagate faster than $c$, no part of the boundary of a 
$D$-dimensional 
lump cannot be spacelike. Therefore, from the covariant conservation laws 
of the form 
\begin{equation}\label{17}
 \partial_{\mu} J^{\mu}= \partial_t J^0 + \nabla {\bf J} = 0 \; ,
\end{equation}
it follows that various quantities are conserved in time, but not 
in space: 
\begin{equation}\label{18}
 \frac{d}{dt} \int_{V} d^{D-1} \! x \: J^0 ({\bf x},t) =0 \; . 
\end{equation}
 
Let us now consider the question why we cannot travel in time. This
question can be answered from several points of view, corresponding to
slightly different definitions of the notion of time travel. First, one
can argue that a time traveler can observe that he arrived at the past
only if he remembers the future, which is extremely improbable. 
The second approach is based
on the consideration of the difference between space and time travel. The  
fact that material objects can travel in both space directions but   
only in one time direction can be stated rigorously as: The trajectory of a
material object ${\bf x}(t)$ is a single-valued function, whereas its inverse
$t({\bf x})$ is not necessarily a single-valued function. To clarify this, let
us consider a $1+1$ dimensional example of a trajectory which would
correspond to the time travel in that sense: $t(x)=-x^2$. This can be
viewed as an object traveling first in the positive time direction,
but at $t=0$ it starts to travel in the negative time direction. However,
this is how it really would look like for an independent observer: Two
{\em identical} objects (which is rather improbable by itself if these are
not
two elementary particles) approach each other, they finally collide at
$t=0$ and then disappear for $t >0$, thus violating the conservation laws.
In other words, objects can travel in both space directions, but only in
one time direction because they are localized in space and thus conserved in
time.

The third approach to the time travel, based on the possibility that the
Universe can possess topology or a metric tensor which admits closed 
timelike curves, is the subject of many current theoretical investigations. 
One of the most important contributions against the 
time travel is given in \cite{hawk3}, where it is 
argued that various conditions (topological defects and metric   
tensors which do not possess the $(1,D-1)$ signature everywhere) 
needed for various 
mechanisms of time travel cannot be realized in practice, essentially
because their realizations require infinite energy. 
However, in my opinion,
the strongest argument against the time travel, discussed also in 
\cite{hawk3} and particularly clearly in \cite{krasnikov}, is
the consistency requirement: for any space-time point $x$, all physical
fields $\varphi_a (x)$ must be {\em uniquely} determined. The 
consistency in the Cauchy-problem approach is automatically provided 
by Axioms \ref{ax1} and \ref{ax4}. 
The time travel based on metric 
tensors which do not possess the $(1,D-1)$ signature everywhere is also 
excluded by Axiom \ref{ax2}. In \cite{hawk3} it is
also
shortly discussed the possibility of time travel if it is possible to
travel in space faster than light, but we have already excluded the possibility   
of traveling faster then light. 

Now a few notes on the different roles of time and space in the 
Hamiltonian formalism. 
Historically, the Hamiltonian formalism was first developed for pointlike    
particles, i.e., for the objects which are strictly localized in space and
exist for all times. This is the difference between the role of
space and time already at the 
kinematic level. Thus, it is not strange that particle mechanics has a
formulation, such as the Hamiltonian formalism, which treats space and time
in different ways.

However, such an argument cannot be directly applied to field theories.
The Hamiltonian formulation of them was probably partly influenced by our
intuitive notion of time, which is the consequence of the time arrow, leading
to the intuitive picture that dynamics is something that changes
with {\em time}. This leads to the notion of ``degree of freedom" as a
real variable which can (at the kinematic level) possess arbitrary
dependence on time. Thus, the set of all degrees
of freedom of a real scalar field is given by all space points ${\bf x}$,
not by all space-time points $({\bf x},t)$. Dynamics, i.e., an equation
of motion, is something that determines the actual time dependence. In the
Hamiltonian approach to field theory, dynamics is given by the Hamiltonian
density ${\cal H}={\cal H}(\phi({\bf x}), \pi({\bf x}))$. The Poisson brackets
among functions of $\phi({\bf x})$ and $\pi({\bf x})$ are actually
equal-time Poisson brackets \cite{nik2}. They can be
viewed
as Poisson brackets among initial conditions. Thus, the phase space is        
space of all initial conditions. In the spirit of the axioms of Section 3,
it
is most natural to consider {\em the degrees of freedom as variables
which can be arbitrarily chosen on ${\cal M}_B$} (except that they must be
essentially regular and ordered). Such a viewpoint will be exploited for
the formulation of the canonical quantum theory.
 
I want to emphasize that the canonical formalism 
in {\em classical} field theory is only a convenience of calculation. 
Nothing is really lost if one does not at all introduce Hamiltonians and
Poisson brackets, but rather uses only Lagrangians and corresponding 
manifestly covariant equations of motion. The existence of the Hamiltonian
formalism in classical field theory still does not mean that space and 
time take different roles. For example, one could also formulate a        
variant of the canonical formalism  
in which the $x^1$ coordinate takes a special 
role, by introducing the Legendre transformation 
\begin{equation}\label{1c1}
 {\cal H}^{(1)}=\pi^{(1)}(x) \frac{\partial\phi(x)}{\partial x^1} -
 {\cal L} \ ,
\end{equation}
where
\begin{equation}\label{1c2}
 \pi^{(1)}(x) = \frac{\partial {\cal L}}{\partial (\partial_1 \phi(x))} \; .
\end{equation}
(Note that (\ref{1c1}) is equal to  
$\Theta_1^1$ in (\ref{14})).    
In particular, this would lead to a new kind of Poisson brackets which
would be interpreted as equal-$x^1$ Poisson brackets.
  
At the end of this section let me give a few notes on theories with
constraints. The constraints appear in 
the Lagrangians which are invariant with respect to some local gauge 
transformations \cite{padm}. For such systems, some of the equations 
of motion are interpreted as constraint equations, which can be understood 
as constraints to the initial condition. Thus the initial condition is not
arbitrary, i.e., the number of
fields which can be arbitrarily fixed on the initial Cauchy surface 
is smaller than
it seems at first sight. Axiom \ref{ax5} refers to these physical
degrees of freedom, which are 
actually the fields for which the initial condition can be
arbitrarily chosen (this refers to their initial time derivatives too),
whereas the initial values of other fields are determined via the constraint
equations. In order to provide a well-posed Cauchy problem, 
some additional gauge conditions must be 
chosen before determining the time evolution. 

\section{The differences between space and time in quantum physics}
\label{SEC5}

In this section I discuss the origin of the differences between space and
time in quantum physics. The connection with
classical physics is the most manifest in the Heisenberg picture, which 
I use to formulate the quantization as a boundary condition for field
operators. As in Sec.~\ref{SEC4}, for the sake of completeness and clarity,  
I also review some already known results.
The main new idea of this section is a suggestion that quantum
physics cannot remove, in a satisfactory way, singularities of the
corresponding classical theory, so we need to modify the {\em classical} 
theory in order to remove the singularities.  

The Heisenberg-picture quantization is based on equal-time commutators 
among canonical coordinates and conjugated momenta, which gives 
different roles to time and space. One could wonder whether we  
can use the equal-space Poisson brackets resulting from
the formalism based on (\ref{1c1}) to propose the corresponding
equal-space commutation relations, without changing the physical content
of the resulting theory. The answer is {\em no}, owing to the fact
that
the Poisson brackets are {\em defined} to be what they are, while the
corresponding commutation relations are {\em postulated}. In other  
words, introduction of the Poisson brackets does not change the physics,
while introduction of the commutation relations does change the physics. 
Thus the difference between space and time in quantum physics is 
even deeper than in classical physics.   

Let me stress some other important facts about the Heisenberg-picture 
quantization. 
The ``general" solution of the equation of motion 
for a free real scalar field, which is usually used, is  
\begin{equation}\label{30}
 \phi(x)=\int\frac{d^{D-1} k}{(2\pi)^{D-1} 2\omega}
 [a(k)e^{-ik\cdot x} + a^{\dagger}(k)e^{ik\cdot x}] \; ,
\end{equation}
where $\omega = \sqrt{{\bf k}^2 + m^2}$, and integration is performed over all 
real vectors ${\bf k}$. 
Let us emphasize once again that this is not really the general solution, 
because there are also other solutions connected with imaginary
$\omega$ and ${\bf k}$. 
However, this {\em is} the general solution if we restrict
ourselves to the solutions which are consistent with Axioms \ref{ax5}
and \ref{ax6}. A more general solution would lead to different 
physical results. In particular, 
fields would not commute for spacelike separations. 

There is one more important property of the operator $\hat{\phi}({\bf x},t)$ 
and its corresponding Hilbert space. For any fixed instant $t=t_0$ and for
any
regular function $\phi({\bf x})$, there is a Hilbert state $|\psi\rangle$
such that 
\begin{equation}\label{34}
 \hat{\phi}({\bf x},t) |\psi\rangle = \phi({\bf x}) |\psi\rangle \; ,
 \;\;\;\;\;\; {\rm for} \;\;\; t=t_0 \; .
\end{equation}   
A similar statement is true for the operator $\hat{\pi}({\bf x},t)$.
However, similar statements are not true if the roles of time coordinate
and one of the space coordinates are exchanged. 

This fact leads to an important 
additional physical motivation for Axiom \ref{ax5}. 
This axiom essentially says that singular field configurations can never form
in a proper classical theory. We know very well that Einstein's theory   
of gravity does not possess this property, because it leads to cosmological
and black-hole-like singularities. Almost everyone agrees that
singularities do not really exist in the real world. However, there is a
wide belief that {\em quantum} theory of gravity, when found one day, could
remove such pathologies, even if the corresponding classical theory does  
possess these pathologies. 
I want to argue that quantum physics cannot remove, in a satisfactory way,
the singularities of the corresponding classical theory; 
we should rather modify the theory of gravity   
for strong fields already at the classical level. 

For this reason, I consider
a simple example: a particle moving in a spherically symmetric potential
$V(r)$, such that $V(\infty)=0$ and $V(0)=-\infty$. Classically, the 
particle can fall into the potential well, thus reaching infinite
kinetic energy (but finite total energy, which is the constant of the motion
and is the sum of the kinetic and the potential energy). It is often said 
that quantum physics prevents such pathological behavior because it    
prevents the particle falling into the center of the potential well. But is this
really true? The Schr\"{o}dinger equation gives a set of eigenfunctions
of the Hamiltonian $\{ \Psi_n ({\bf x};t)=\psi_n ({\bf x}) e^{-iE_n t} \}$,
which serves as a basis for the general solution of the Schr\"{o}dinger
equation. This means that the particle can be found  
everywhere, including the singular point $r=0$. The set of functions 
$\{\psi_n ({\bf x})\}$ is complete, which means, in particular, that the
wave function 
at some particular instant can be proportional to $\delta^{D-1}({\bf x})$,
or to $e^{i{\bf p}\cdot{\bf x}}$ for any
particular ${\bf p}$, which are eigenstates of the operators 
$\hat{{\bf x}}$ and $\hat{{\bf p}}$, respectively. 
In other words, the particle can attain any position or 
any momentum. The only restriction is that these two quantities are not
mutually independent, because the corresponding operators do not commute. 
In the language of energy, the particle can possess any mean potential   
energy or any mean kinetic energy, including the infinite one.
The only restriction is that it cannot possess a mean total energy 
smaller than the ground-state energy $E_0$. And this is not a much better
situation than in classical physics, because in a typical physical
classical situation we do not expect a minus infinite total energy either. 

However, why is an atom still stable? This is because the probability
density for states with a fixed energy $P_n({\bf x})=|\psi_n ({\bf x})|^2$ 
does not depend on time, so lasts forever. On the other hand, if the  
wave function is proportional, for example, to $\delta^{D-1}({\bf x})$ at
some
instant $t$, then it is a state which possesses components of many admissible
energies. Thus $P({\bf x})$ changes with time, being strictly localized
only at one particular instant $t$. Thus we have much better chances    
to find the particle in a state with a fixed energy.

Similarly, if the classical theory of gravity possesses a singular solution
$g_{\mu\nu}({\bf x})$ for some instant $t$,
then we must expect that in the corresponding 
quantum theory there exists a state $|\psi\rangle$ which corresponds to
this solution at some instant $t$. The best we can expect is that we shall
never observe such a state because it lasts too short. However, I    
believe that singular states should not exist at all, so I require that   
singularities should not appear even in classical physics.
 
Now we are finally ready to propose an axiom for the quantization of fields. 
It must explain, rather than postulate, 
why time has a special role in quantization and why in (\ref{30}) 
we take only real $\omega$ and ${\bf k}$. We do not know
how to canonically quantize theories with higher than second derivatives
in the equations of motion, but it seems that such theories cannot be
consistent with Axiom \ref{ax5}, as I discuss in  
Section 6. Thus I assume that all fields that can be arbitrarily chosen
(except that they must
be essentially regular and ordered) on ${\cal M}_B$, can be divided into a
set
of fields $\{\varphi_a \}$ and conjugate momentum fields $\{\pi_a \}$,
where
\begin{equation}\label{35}
 \pi_a = \frac{\partial {\cal L}}{\partial (\partial_t \varphi_a)}
\end{equation}
and $t$ is the coordinate defined as in Sec.~\ref{SEC4}.
Having all this in mind, I propose:  
\begin{axiom}\label{ax8} 
 Let ${\bf x},{\bf x}' \in {\cal M}_B$.
 All fields $\{ \varphi_a \}$ and $\{ \pi_a \}$ are quantized in such a way
 that 
 \begin{eqnarray}\label{37}
  & [\hat{\varphi}_a ({\bf x},0),\hat{\pi}_b ({\bf x}',0)]_{\pm} 
  =i\delta_{ab}\delta^{D-1}({\bf x}-{\bf x}') \; , & \nonumber \\    
  & [\hat{\varphi}_a ({\bf x},0),\hat{\varphi}_b ({\bf x}',0)]_{\pm}=
    [\hat{\pi}_a ({\bf x},0),\hat{\pi}_b ({\bf x}',0)]_{\pm}=0 \; . &
 \end{eqnarray} 
 Furthermore, the field operators $\{ \hat{\varphi}_a (x) \}$ and 
 $\{ \hat{\pi}_a (x) \}$ satisfy classical equations of motion and  
 they are quantized in such a way that for given functions  
 $\varphi_a({\bf x})$ and $\pi_a({\bf x})$ there exist states
 $|\psi_{\varphi_a}\rangle$ and $|\psi_{\pi_a}\rangle$ such that
 \begin{eqnarray}\label{38}
  & \hat{\varphi}_a ({\bf x},0) |\psi_{\varphi_a}\rangle =
    \varphi_a({\bf x}) |\psi_{\varphi_a}\rangle \; , & \nonumber \\
  & \hat{\pi}_a ({\bf x},0) |\psi_{\pi_a}\rangle = 
    \pi_a({\bf x}) |\psi_{\pi_a}\rangle \; , &
 \end{eqnarray}
 if and only if $\varphi_a({\bf x})$ and $\pi_a({\bf x})$ are
 essentially regular functions.
\end{axiom} 
It is, of course, understood that we use anti-commutators if both fields
possess half-integer spin and commutators otherwise. 
For fermion degrees, $\varphi_a({\bf x})$ and $\pi_a({\bf x})$ 
are products of a complex essentially regular function and a 
Grassmann number. Since this quantization 
is canonical, it is not manifestly covariant. However, we expect
that covariance is 
preserved because the field operators satisfy the covariant equations        
of motion. This can be explicitly proved for free fields and on the  
perturbative level for fields in interaction, but I shall not consider
these rather technical problems. One of the most important consequences of
covariance is that the statements of Axiom \ref{ax8} are valid not
only for $t=0$, but also for all other times, obtained by time evolution or
coordinate transformation.  

Axiom \ref{ax8} can be understood as an initial condition for the 
field operators. It proposes that we have to quantize those
classical variables which can be arbitrarily chosen on ${\cal M}_B$. It 
can also be viewed as an explanation why in the quantum theory of 
particles (i.e., first quantization) there is an $\hat{{\bf x}}$-operator, 
but there is no $\hat{t}$-operator.  

An important ingredient of Axiom \ref{ax8} is that it
proposes that {\em only physical degrees of freedom should be quantized}. 
This is extremely important for quantum gravity, because the quantum theory 
of gravity in which both physical and nonphysical degrees are 
quantized is not equivalent to the theory in which only physical degrees 
are quantized. The quantization of the physical degrees only is also one 
of the ways how to solve the problem of time in quantum gravity
\cite{isham}. 

It is also important to note that the consistency of the canonical quantization 
requires the topology of 
spacetime to be $\Sigma\times\mathbb{R}$, which is provided by  
Axiom \ref{ax1}.   

It is straightforward  to convert operators and states from the 
Heisenberg to the Schr\"{o}dinger picture. This leads to the 
functional Schr\"{o}dinger equation, which determines the wave functional 
$\Psi [\phi({\bf x});t)$, being a functional with respect to $\phi({\bf x})$ 
and a function with respect to $t$.   
Both the Heisenberg and the Schr\"{o}dinger picture of quantum field theory
manifestly express the fact that space and time are not treated in the same
way. On the other hand, it is usually stated that 
the functional-integral formulation 
is manifestly covariant, which might seem to be in
contradiction with the fact that space and time have different roles 
in quantization. However, space and time have different roles 
even in the functional-integral formulation, because it is given 
by 
\begin{eqnarray}\label{42}
 & \langle\phi_f({\bf x}),t_f | \phi_i({\bf x}),t_i\rangle  =   
 \displaystyle\int [d\phi({\bf x},t)] [d\pi({\bf x},t)] \times & \nonumber
\\
 & \exp \left\{
 i\displaystyle\int_{t_i}^{t_f}\! dt\! \displaystyle\int\! d^{D-1}x   
 \left[\pi({\bf x},t) \dot{\phi}({\bf x},t)-
 {\cal H}\left(\phi({\bf x},t),\nabla\phi({\bf x},t),\pi({\bf
x},t)\right)\right]
 \right\} \; , & 
\end{eqnarray}
where $|\phi({\bf x}),t\rangle \equiv \Psi [\phi({\bf x});t)$. 
The left-hand side obviously gives different roles to space and time. This 
is manifested on the right-hand side in the fact that the functional integral
is not performed over all functions $\phi({\bf x},t)$, but only over
functions 
which satisfy     
\begin{equation}\label{43}
 \phi({\bf x},t_f)=\phi_f({\bf x}) \; , \;\;\;\;\;
 \phi({\bf x},t_i)=\phi_i({\bf x}) \; .
\end{equation}
Furthermore, $t$ takes values from the finite interval $t\in [t_i,t_f]$,
while
${\bf x}$ takes values from the infinite interval ${\bf x}\in \mathbb{R}^{D-1}$.
At the end, the sub-integral function
$\pi\dot{\phi}-{\cal H}(\phi,\nabla\phi,\pi)$ is
not Lorentz invariant. The invariant form is obtained only when  
the $\pi$-dependence is integrated out, the vacuum-to-vacuum amplitude is
considered, and $t_i\rightarrow -\infty$, $t_f\rightarrow \infty$; 
\begin{equation}\label{45}
 \langle\phi_f({\bf x})=0,t_f\rightarrow \infty |
 \phi_i({\bf x})=0,t_i\rightarrow -\infty\rangle \equiv Z
 = \int [d\phi(x)]
 \exp\left\{ i\int d^D x \: {\cal L}(\phi(x),\partial_{\mu}\phi(x))\right\}
 \; .
\end{equation}
However, the left-hand side still treats space and time in different ways
and the functional integral on the right-hand side is still restricted to    
functions which satisfy (\ref{43}). The general expression (\ref{42}) 
is equivalent to the Schr\"{o}dinger equation, while the Lorentz-invariant 
expression (\ref{45}) is only a special case, from which 
the Schr\"{o}dinger equation cannot be derived. 

It is also important to note that in (\ref{42}), for a given space-time
point $({\bf x},\: t\neq t_i ,t_f)$, the integration is performed
over {\em all} possible finite real values of 
$\phi$ and $\pi$. This is the direct consequence of the fact that for any
regular functions $\phi({\bf x})$, $\pi({\bf x})$ and for any $t$ there
exist
states such that these functions are eigen-values of the corresponding
field operators. This means that in theories with constraints the functional
integral is performed only over the physical degrees of freedom, which is 
important for quantum gravity. Note also that,
according to my axioms, in the case of quantum gravity there is no sum
over topologies; only the global $\mathbb{R}^D$ topology is included. 
     
Let us now discuss the meaning of the discussion presented in         
Sec.~\ref{SEC4} from the point of view of quantum field theory. Although the    
whole Sec.~\ref{SEC4} refers to the classical field theory, all arguments are 
correct at the macroscopic level, because we know that classical theory
is a good approximation at the macroscopic level. In particular, the
law of disorder increasing, as a statistical law, is valid only on
the macroscopic level. On the other hand, 
there are arguments that quantum mechanics
possesses the intrinsic, fundamental time arrow, connected with the ``fact"
that wave functions collapses. However, excellent
arguments against such conclusions are given in \cite{unruh}. 
There are also arguments, based on
the considerations of the wave function of the Universe, that entropy would
start to decrease when the Universe starts to contract.
It is remarkable to note that Hawking was the first that came to such a
conclusion \cite{hawk5}, but later he corrected himself \cite{hawk2},
claiming that his conclusions were based on certain misinterpretations.
A general discussion on various misunderstandings of the time arrow
is given in \cite{price}. It seems to me that all conclusions
made by some authors about the
different status of the time arrow in classical and quantum physics, if
not incorrect, are at least interpretation dependent, because there
are various interpretations of quantum mechanics and no one knows yet which
is the correct one. The origin of the collapse of the wave function
is still not understood. My personal belief is that quantum mechanics is 
just some effective, incomplete theory, while the underlying more 
fundamental theory is some deterministic nonlocal hidden variable theory,
which obeys some laws not very different from the axioms of Section 3. 
Actually, it is very likely that only Axiom \ref{ax3} should be modified.
For example, in the de Broglie--Bohm interpretation of quantum field
theory \cite{holl}, the classical equations of motion are modified by 
adding an external force proportional to $\hbar^2$, in which 
fields are integrated over space, but not over time. This term breaks 
Lorentz covariance and locality, but 
the resulting theory still possesses a 
well-posed initial-value problem. In this interpretation, Lorentz   
covariance and locality are statistical effects, which are the only ones
measured in present experiments. 
        
Having in mind the remarks of the last paragraph, we may conclude 
that quantum mechanics probably does not change the origin of the    
time arrow. 

\section{Do our theories satisfy Axiom \protect\ref{ax5}? }
\label{SEC6}

We have argued that equations of motion must obey some properties, such as    
hyperbolicity and boundedness of energy from below, in order to
satisfy Axiom \ref{ax5}. However, nothing provides that these properties
are enough. We have to check whether our theories really satisfy this   
axiom, and if they do not, whether they can be modified in such a 
way as to still agree with present observations.  
I give only some qualitative discussion of this, without 
intention to be rigorous. 

Let us start from electrodynamics. Electromagnetic fields and charges
obey
Lenz's law, which essentially states that any change tends to be  
canceled. This speaks in favor of satisfying Axiom \ref{ax5}.
One could argue that classical electrodynamics has problems with
infinities connected with pointlike charges. However, one should not 
forget that we are considering a {\em field} theory of charges, i.e., 
continuous distributions of charge. Because of Axiom \ref{ax6},  
there are no initial infinite charge densities and thus there are no initial    
pointlike charges. Since the force among charges of the same sign   
is repulsive, pointlike charges will never form. Of course, both   
classical and quantum electrodynamics still cannot 
determine the size of the electron and its electromagnetic mass. 
But the important thing is that classical
electrodynamics does not {\em predict} the singularities of this kind. 

However, it seems that classical electrodynamics
can still lead to some divergences under very specific initial conditions.
For example, one can consider a
free electromagnetic wave which is exactly spherically symmetric and
moves toward the center of the sphere. This will result in an infinite
energy-density in the center when the wave comes there. However, the
Lagrangian of
electrodynamics is certainly not correct for very strong fields,
so it is very likely that formation of such infinities
is prevented on high energy scales, by some 
yet unknown interactions. Similar discussion can be done for all other
non-gravitational interactions. 

The inconsistency of Einstein's theory of gravity with Axiom \ref{ax5}
is more obvious than that of other theories, because it is shown by Hawking and
Penrose \cite{hawpen} that singularities will develop under very general
initial conditions in Einstein's classical theory of gravity. This is
one of
the motivations to find an alternative theory of gravitation. The status of
singularities in various alternative theories of gravitation is reviewed
in \cite{fuchs}. 

One class of alternative gravity theories are higher derivative theories,
based on addition of higher powers of the curvature tensor to the Lagrangian
of  
Einstein's theory. However, even if these terms can prevent cosmological
and
black-hole-like singularities, their inconsistency with Axiom \ref{ax5}
is even more obvious. It turns out \cite{boulw} that in such theories
the energy is not bounded from below and thus runaway solutions appear.
Similar problems appear in various non-gravitational higher derivative 
theories as well. There is no general theorem which provides that every higher
derivative partial differential equation possesses such problems, but such
problems are found in physically interesting cases. This is why we
usually disregard higher derivative theories. It is also often claimed
that
such theories violate causality. This is because one needs to impose the
boundary conditions at $t\rightarrow \pm\infty$ in order to remove these
runaway solutions. The presence of runaway solutions is also connected to
the violation of Einstein causality, i.e., to non-vanishing (anti)commutators
outside the light-cone. This connection can be easily seen on the example 
of tachyon fields \cite{fein}.    
    
Another class of generalizations of Einstein's theory of gravitation are 
gauge theories of gravity
\cite{ivan}, \cite{gronw}. The most important of them 
is the Einstein-Cartan theory, 
which leads to the existence of torsion. It turns out that singularities
in such theories do not develop under such wide conditions as in 
Einstein's theory, but they can still appear, for example, for spin-less
matter, which does not feel torsion.    

The third class of alternative gravity theories, perhaps most in the
spirit of the philosophy of this article, are bi-metric theories. The main
idea is to separate the metric tensor into two parts 
\begin{equation}\label{46}
 g_{\mu\nu}=\gamma_{\mu\nu} + \Phi_{\mu\nu} \; ,
\end{equation}
where $\gamma_{\mu\nu}$ is a non-dynamical, background metric, while
$\Phi_{\mu\nu}$ is a dynamical field, determined by some differential
equations. Such theories are often called ``field theories of gravitation"
because such theories are the most similar to other field theories,
describing a field in a fixed background metric. In theories of this 
kind it is manifest that topology is not dynamical, but rather fixed by the 
background metric $\gamma_{\mu\nu}$. 

One of the variants of bi-metric theories 
is the theory developed by Logunov and
others
\cite{log}. The motivation for this theory has been criticized \cite{zeld} 
because this theory was motivated by some incorrect criticism of Einstein's
theory of gravity. However, Logunov's theory itself is self-consistent and
still
possesses some advantages with respect to Einstein's theory. The background 
metric in this theory is flat Minkowski metric
$\gamma_{\mu\nu}=\eta_{\mu\nu}$.
The metric $g_{\mu\nu}$ of a spherically symmetric object with a mass
$M$ takes the same form as a Schwarzschild solution for $r\gg 2MG$. 
However, a small mass $m$ is attributed 
to the gravitational field $\Phi_{\mu\nu}$, whose effect is that
the gravitational force becomes repulsive for strong fields, thus preventing 
black-hole and cosmological singularities.\footnote{There are
arguments that even a small mass cannot
be attributed to a graviton because it would significantly deviate from
experiments even in a small mass limit \cite{dam}.
However, these arguments are applicable only to Einstein's theory of  
gravity,
not to any theory of gravity. The effects of a small enough graviton mass
in Logunov's theory are in agreement with experiments \cite{chug}.}     
A homogeneous and isotropic 
universe is infinite in space, exists for an infinitely long time and
oscillates. Thus it 
seems that this theory satisfies Axiom \ref{ax5} and is manifestly in 
agreement with Axiom \ref{ax1}. However, I am far from saying that this
is the right theory. For example, the corresponding quantum variant 
is certainly not renormalizable, essentially for the same reasons as 
Einstein's theory, because the same dimensional coupling constant $G$
appears in the Lagrangian. 
I am just arguing that this theory could be closer to the right theory
which we do not know yet. 

Let us discuss at the end why there are no fields with spin higher than 2.
Their status is similar to the theories with derivatives higher than second;
there is no general theorem, but the simplest theories constructed, for
example, in \cite{spin}, possess some pathologies. They violate Einstein
causality, i.e., they propagate faster than light and (anti)commutators  
do not vanish outside the light-cone. They also violate Cauchy
causality, i.e., the Cauchy problem is not well posed. 
The axioms of Sec.~\ref{SEC3} assume, of course, that the
Cauchy problem must be well posed. The general relation 
between Einstein and Cauchy causality is discussed in \cite{keis}.

\section{Discussion}      
\label{SEC7}

As discussed already, only Axioms \ref{ax1} and \ref{ax5} are really  
novel principles, in the
sense that they differ from the conventional point of view and can be
tested, at least in principle. Here I want to discuss whether these  
axioms can be rejected or weakened and what consequences of
this would be.
I shall also give a few comments on the dimensionality of space.

Axiom \ref{ax5} essentially says that for any finite everywhere initial
condition the solution is also finite everywhere. This axiom explains the
hyperbolicity, i.e., the $(1,D-1)$ signature of the metric. It also explains   
the absence of tachyons, the positivity of energy, and other related properties  
of nature. However, from these properties Axiom \ref{ax5} certainly
cannot be derived. 

First, there is a possibility that infinities do 
exist, but almost no one believes that. 

A much more probable possibility 
is that nature somehow chooses only those initial conditions that will
not lead to infinities. However, such a principle is quite unaesthetic;
Axiom \ref{ax6} seems much simpler and more natural than this one.

The best alternative is probably the assumption that singularities can occur
in classical physics as long as quantum physics prevents them. However, as
we have already discussed, quantum physics cannot prevent the existence of
states which correspond to the singular behavior at some particular instant
of time. The best we can expect from quantum physics is that it is
practically impossible to observe such states. One can be satisfied with 
this, but Axiom \ref{ax5}, together with Axiom \ref{ax8}, is more 
satisfying, because it provides that singular states do not exist at all. 

It is difficult to test Axiom \ref{ax5} experimentally, because we cannot   
measure infinities. However, finding tachyons, for example, would be a strong
argument against this axiom. But this would also violate some widely  
accepted principles, such as Einstein causality.

A more serious question is whether the topology is really a non-dynamical
entity, as proposed in Axiom \ref{ax1}. I want to emphasize once again
that the topology is a more fundamental concept than the metric tensor, 
in the sense
that the former can be defined without the latter. And the Einstein equation
is manifestly a theory of the metric tensor, not of the topology. If the Einstein  
equation is treated as a Cauchy problem, for example, by numerical
computation, the manifold of space-time points and its topology {\em must}
be
defined before any computation of the metric tensor is performed. 
If the Cauchy problem is well posed, then the space topology 
cannot change during the time evolution \cite{konst}. The fact
that some solutions of the Einstein equation correspond to some 
topologies still does not mean that the Einstein equation describes the 
topology; it merely means that the solution must be consistent with a 
given topology. Moreover, the metric tensor does not even uniquely determine    
the topology. For example, the flat metric $\eta_{\mu\nu}$ does not
necessarily imply that the corresponding manifold is infinite; it can also
correspond to a torus or a cylinder. A similar statement is true for any
other differential equation; if the solution $\phi(x)$ satisfies some
periodicity
conditions, we still do not know whether this solution corresponds to a
closed
or an infinite manifold (i.e., set of points $\{ x \}$). If the Einstein equation
can say anything at all about the topology, it can do that only indirectly. 
At least, this is so in classical gravity. Can quantum gravity change  
this? The set of space-time points and its topology is certainly a 
non-dynamical entity in all non-gravitational theories, both classical
and quantum. We just argued 
that this is also so in classical gravity. So I really do not see 
why quantum gravity would change this, at least if quantum gravity is based
on the quantization of some classical theory of the {\em metric tensor}, such as
Einstein's theory. This can be seen most explicitly in the Heisenberg  
picture; one writes the general solution of classical equations
consistent with a given topology and then just promotes all free parameters  
to the operators (assuming that one can solve technical problems connected
with
this). This can also be seen from the kinematics of the wave function
$\Psi [g_{\mu\nu}({\bf x});t)$; since it is a functional of 
$g_{\mu\nu}({\bf x})$, it can be defined only if the set of points $\{ {\bf
x}
\}$ is previously well defined.  
However, even if one proposes that various topologies must be allowed in 
quantum theory, for example, by summing over topologies in a functional integral
(although it is not clear what would then stay on the left-hand side of the
analog of (\ref{42}) and what the analog of the condition
(\ref{43}) would be),
then one would expect that the sum over various signatures, or even
dimensionalities of spacetime should be performed as well, 
because the Einstein equation itself does not fix them either. However, for
some
reason, such a possibility is not usually considered. 
The sum over dimensionalities in quantum gravity would 
imply that even in nongravitational quantum theories the sum over dimensionalities
should be performed.  

If the topology must be fixed, as I just argued, the next question is what 
{\em is} the topology of the Universe? In order to Cauchy problem be 
well posed and canonical quantization possible, it is necessary that 
the topology is of the form $\Sigma \times\mathbb{R} $. There are no 
inconsistencies (as far as I know) for any choice of a connected,  
orientable 
$(D-1)$-dimensional manifold $\Sigma$ 
without a boundary. Closed $\Sigma$'s would still allow 
only oscillatory solutions, such as $e^{i{\bf k}\cdot {\bf x}}$, no longer
by the finiteness requirement, but rather by the periodicity requirement.
However, the choice 
$\Sigma=\mathbb{R}^{D-1}$ is the simplest and 
leads to the highest degree of symmetry 
between space and time. Thus Axiom \ref{ax1} seems to be
very natural. Of course, this axiom still allows effective closed 
topologies by an ``accident", if solutions of the equations of motion satisfy
some
periodicity conditions. If all fields (and wave functions) satisfy 
appropriate periodicity conditions, no observation can distinguish the
``really" closed universe from the periodic one. 

At the end, let me make a few comments on the dimensionality of space.   
The axioms of this 
article certainly cannot explain why space is 3-dimensional.
The answer to this question should be searched elsewhere. For example,
superstring theory predicts that $D=10$. It still cannot explain  
why 6 coordinates are compactified. But if they are, 
this is not in contradiction with Axiom \ref{ax1}, 
as I just discussed. Some types of effective
compactifications, such as torus $T^6$, are still possible. 

There are interesting attempts to explain why space is 3-dimensional based 
on certain anthropic considerations \cite{teg}. However, such arguments
do not seem too convincing to me. 

\section{Conclusion}
\label{SEC8}

All differences between the role of space and time in nature
can be explained by proposing a set of principles in which none of
the space-time
coordinates has an {\it a priori} special role. The essence of my approach is a
proposal that all dynamical field equations {\em must} be treated as a
Cauchy problem. This requires that the topology of spacetime must be fixed at
the predynamical level. 
Various choices of topology of the form $\Sigma\times \mathbb{R}$
are admissible, but the choice $\mathbb{R}^D$ is the most natural and is the only
one that does not give an {\it a priori} special role to any coordinate. The
hyperbolicity, i.e., $(1,D-1)$ signature of the metric, can be explained by
proposing that {\em any} boundary condition that is finite everywhere
must lead to the solution which is also finite everywhere. It also  
explains the 
boundedness of energy from below, the absence of tachyons, and other related
properties of nature. It is quite likely that this principle must be
realized in nature because it automatically prevents all kind 
of physical singularities. The time arrow can be explained by proposing that
the boundary condition is ordered, rather than random. 
The quantization can be considered as a boundary condition for the field
operators. It appears natural to quantize the physical degrees of freedom
only. This, together with the
treatment of spacetime as a non-dynamical background, resolves a lot of
conceptual problems in classical and quantum gravity, including the
problem of time in quantum gravity.  

It was no intention of this article to be mathematically rigorous. The 
main intention was to provide a complete conceptual understanding.   
A more rigorous
treatment, as well as many technical details of some questions 
considered here, can be found in references cited. I hope that
future investigations will also put all other ideas of this article into 
a more rigorous framework.  

\section*{Acknowledgment}

This work was supported by the Ministry of Science of the Republic of Croatia.

\end{document}